# Nonlinear control of photonic higher-order topological bound states in the continuum

Zhichan Hu[1†], Domenico Bongiovanni[1,2†], Dario Jukić[3], Ema Jajtić[4], Shiqi Xia[1], Daohong Song[1,5], Jingjun Xu[1,5], Roberto Morandotti[2,5], Hrvoje Buljan[1,4*], and Zhigang Chen[1,6,7*]

[1] *The MOE Key Laboratory of Weak-Light Nonlinear Photonics, TEDA Applied Physics Institute and School of Physics, Nankai University, Tianjin 300457, China*
[2]*INRS-EMT, 1650 Blvd. Lionel-Boulet, Varennes, Quebec J3X 1S2, Canada*
[3]*Faculty of Civil Engineering, University of Zagreb, A. Kačića Miošića 26, 10000 Zagreb, Croatia*
[4]*Department of Physics, Faculty of Science, University of Zagreb, Bijenička c. 32, 10000 Zagreb, Croatia*
[5]*Institute of Fundamental and Frontier Sciences, University of Electronic Science and Technology of China, Chengdu 610054, China*
[6]*Collaborative Innovation Center of Extreme Optics, Shanxi University, Taiyuan, Shanxi 030006, China*
[7]*Department of Physics and Astronomy, San Francisco State University, San Francisco, California 94132, USA*
[†]*These authors contributed equally to this work*
*[\*hbuljan@phy.hr](mailto:hbuljan@phy.hr); [\*zgchen@nankai.edu.cn](mailto:zgchen@nankai.edu.cn)*

# Abstract:

Higher-order topological insulators (HOTIs) are recently discovered topological phases, possessing symmetry-protected corner states with fractional charges. An unexpected connection between these states and the seemingly unrelated phenomenon of bound states in the continuum (BICs) was recently unveiled. When nonlinearity is added to a HOTI system, a number of fundamentally important questions arise. For example, how does nonlinearity couple higher-order topological BICs with the rest of the system, including continuum states? In fact, thus far BICs in nonlinear HOTIs have remained unexplored. Here, we demonstrate the interplay of nonlinearity, higher-order topology, and BICs in a photonic platform. We observe topological corner states which, serendipitously, are also BICs in a laser-written second-order topological lattice. We further demonstrate nonlinear coupling with edge states at a low nonlinearity, transitioning to solitons at a high nonlinearity. Theoretically, we calculate the analog of the Zak phase in the nonlinear regime, illustrating that a topological BIC can be actively tuned by both focusing and defocusing nonlinearities. Our studies are applicable to other nonlinear HOTI systems, with promising applications in emerging topology-driven devices.

**Introduction**

Over the past decade, topological insulators have attracted tremendous attention across many disciplines of natural sciences[1,2] including photonics[3]. One of their most appealing features is the topologically protected edge states immune to scattering at defects and disorders[4-6]. A few years ago, a novel class of higher-order topological insulators (HOTIs) was predicted[7-9] and experimentally observed[10-12]. While a standard topological insulator usually obeys the bulk-boundary correspondence principle (with edge states one dimension lower than the bulk), the HOTIs typically support zero-dimensional corner states regardless of their physical dimension, or more generally, ($d$-$n$)-dimensional states at the boundaries of $d$-dimensional lattices with $n$ no less than $2^{3,13}$. This is associated with topologically quantized quadrupole and higher-order electric moments in electronic systems[7]. The discovery of HOTIs broadened the concept of the symmetry-protected topological phase and our common understanding of traditional topological insulators, which has thus launched an avalanche of research ventures on HOTIs in a variety of fields, including condensed matter physics, electric circuits, phononic systems, acoustics, and photonics[7,10-12,14-31]. In terms of fundamental interest, HOTIs are attractive because they are related to many intriguing phenomena such as higher-order band topology in twisted Moiré superlattices[32], topological lattice disclinations[33], Majorana bound states[34] and their nontrivial braiding[35]. Towards applications, they have been touted and tested for robust photonic crystal nanocavities[36] and low-threshold topological corner state lasing[28,37].

Combining topology and nonlinearity leads to a number of fundamental questions, some of which have been addressed in the study of first-order nonlinear topological photonic systems[38], including for example nonlinear topological solitons and edge states, nonlinearity-induced topological phase transitions, topological nonlinear frequency conversion, and nonlinear tuning of non-Hermitian topological states[39-46]. However, thus far, all of the studies on HOTIs have mainly been restricted to the linear regime, and only recently it became clear that unexpected phenomena may arise when nonlinearity is taken into account in HOTI systems[47-49], with experiments implemented already in nonlinear electric circuits[47] and photonic structures[50-52].

In a few recent papers, an intriguing connection between HOTIs and another widely studied phenomenon, namely, the bound states in the continuum (BICs), has been explored[21,22,53,54], reinvigorating the interest in BICs and their topological nature as previously established[55-57]. BICs are counter-intuitive localized states with eigenvalues in the continuum of extended states, which may result from versatile mechanisms[56,57]. An exemplary model used for unveiling such a connection is the celebrated two-dimensional (2D) Su-Schrieffer-Heeger (SSH) lattices[58], which possess the second-order localized modes protected by the chiral symmetry and crystalline symmetry[53], as have been observed in a variety of synthetic structures[10-12] including photonic crystals[17,18]. In such HOTIs, the corner-localized states appear right at the center of the eigenvalue spectrum ("zero-energy mode") and are embedded in the continuum band rather than in the gap, in contrast to other types of HOTIs[19,23,59]. Such topological BICs have infinite lifetimes and are fully localized to the corner despite being degenerate with the bulk bands, but they become "leaky" when the required symmetries are broken[53,54]. Nonlinearity can be used to break these symmetries, and thus enable coupling of light into or out of these localized corner states, or performing their braiding, thus making them attractive for potential applications. However, to the best of our knowledge, nonlinear higher-order BICs and their associated dynamics have not been explored so far, in photonics or any other systems.

Here, we establish a nonlinear photonic HOTI platform and explore the role of nonlinearity in higher-order topological BICs (Figs. 1(a, b)). We demonstrate that a low nonlinearity, either self-focusing or -defocusing, can induce coupling between corner states and edge states in a 2D SSH nontrivial lattice, enabling their beating oscillations. Interestingly, the nonlinear coupling of a higher-order BIC to the bulk states is significantly weaker, despite the fact that both chiral and crystalline symmetries are broken by nonlinearity. However, coupling to the bulk readily occurs under the same excitation conditions in a trivial lattice, indicating that the robustness of the weakly nonlinear topological BICs is inherited from the linear system. At a high focusing nonlinearity, a corner state becomes more localized, forming a semi-infinite gap soliton out of the continuum Bloch bands, whereas at a high defocusing nonlinearity it exhibits strong radiation into the edge and bulk (Fig. 1a). Theoretically, we analyze the dynamical evolution of the nonlinear eigenvalue spectrum and validate the robustness of the corner modes in the process of beating with the edge modes driven by a low nonlinearity. This is supported by calculating the bulk polarizations manifesting the topological invariant in the nonlinear regime.

**Results**

The wave dynamics in our HOTI system can be described by the *continuous* nonlinear Schrödinger-like equation (NLSE), typically used for simulating a light field with amplitude $\psi(x,y,z)$ propagating along the longitudinal $z$-direction of the photorefractive photonic lattice[60]:

$$i\frac{\partial \psi}{\partial z} + \frac{1}{2k}\left(\frac{\partial^2 \psi}{\partial x^2} + \frac{\partial^2 \psi}{\partial y^2}\right) - \Delta n \frac{\psi}{1+I_L+I_P} = 0, \quad (1)$$

where $I_L(x,y)$ accounts for the beam intensity for writing the 2D SSH lattice in a nonlinear photorefractive crystal, $I_P$ is the nonlinear contribution of the probe beam, $k$ is the wavenumber in the crystal, $\Delta n$ is the linear refractive index change determined by the bias field $E_0$ and the electro-optic coefficient of the crystal (see Supplementary Material). The induced refractive index change forming the linear photonic lattice depends on the spatial coordinates $x$ and $y$, and it is uniform along the propagation axis as illustrated in Fig. 1a. The nonlinearity can be a self-focusing or – defocusing type, depending on the direction of the bias field relative to the crystalline optical axis, while its strength can be controlled by the bias field and the beam intensity[60,61]. It should be pointed out that for low power probe beams (small $I_P$), the nonlinearity is in the low saturable regime and approximately Kerr-like, then the above NLSE is equivalent to the Gross-Pitaevskii equation that describes interacting atomic Bose-Einstein condensates in the mean-field approximation[2]. As such, even though we used a specific type of optical nonlinearity in our study, the concept and scheme on nonlinear control of HOTI corner modes developed here are expected to hold on other platforms beyond photonics.

The topological features of the 2D SSH lattice are more transparent in the discrete model based on the coupled-mode theory. When the next-nearest-neighbor (NNN) coupling is negligible, Eq. (1) can be approximated with

$$i\frac{\partial \psi_\alpha}{\partial Z} + \sum_\alpha [H_L]_{\alpha,\alpha'}\psi_{\alpha'} + E_0' \frac{\gamma|\psi_\alpha|^2}{1+\gamma|\psi_\alpha|^2}\psi_\alpha = 0, \quad (2)$$

where $\psi_\alpha$ is the complex amplitude of the electric field at the site $\alpha$, $H_L$ is the linear Hamiltonian matrix of the 2D SSH model; its entries $[H_L]_{\alpha,\alpha'}$ are either zero (when $\alpha$ and $\alpha'$ are not neighboring sites), or take the value of either the intra-cell coupling $t$ or the inter-cell coupling $t'$, as illustrated in Fig. 1b. Both the normalized bias field $E_0'$ and the nonlinear coefficient $\gamma$ control the saturable nonlinearity, which corresponds to the nonlinear photorefractive crystal used in the experiment[60,61].

In the linear regime, in full analogy with the 1D case[62,63], the 2D SSH lattices exhibit two distinct Zak phases, which correspond to bulk polarizations. These are the topological invariants for this type of HOTIs[17,18,58], which differentiate the topologically trivial and nontrivial structures (see Methods). They can be tuned by the dimerization parameter $c = t - t'$. In the discrete model, the topological characteristics of the 2D SSH lattices can be seen clearly from the linear eigenvalue spectrum calculated from $H_L \varphi_{L,n} = -\beta_{L,n} \varphi_{L,n}$, as summarized in Figs. 1(c, d), where $\beta_{L,n}$ is the linear spectrum, and $\varphi_{L,n}$ are the corresponding eigenmodes. When the intra-cell coupling is weaker than the inter-cell coupling ($c < 0$), the system is in the topologically nontrivial phase, and the band structure consists of characteristic edge and corner modes (Fig. 1c). In particular, in the middle of the band, there are four degenerated corner modes (Fig. 1d1), corresponding to "zero-energy modes" in the condensed matter language. A typical corner mode structure is shown in the upper inset of Fig. 1d1, which clearly displays the features of the topological corner state (highly localized at the corner with zero amplitude in its nearest-neighbor site but nonzero out-of-phase amplitude in its NNN sites along the edges). Since these corner states are embedded in the continuum of the SSH lattice as well as protected by the $C_{4v}$ and chiral symmetries – they are topological BICs[53,54]. For comparison, the trivial phase, manifested by vanishing polarizations, occurs when the intra-cell coupling is stronger than the inter-cell coupling ($c > 0$), where there are two mini-gaps formed only by the bulk modes (Fig. 1d3). When the coupling strength is uniform across the whole lattice ($c = 0$), it turns to a trivial square lattice with a gapless spectrum (Fig. 1d2), which sets apart the topologically nontrivial and trivial regimes. Representative edge and bulk modes are also displayed in the insets of Fig. 1d.

In the nonlinear regime, the nonlinear eigenvalues are calculated from $H_{NL} \varphi_{NL,n} = -\beta_{NL,n}(Z) \varphi_{NL,n}(Z)$ where the nonlinear Hamiltonian $H_{NL} = H_L + V_{NL}$ contains both the linear part and the nonlinear potential corresponding to the third term in Eq. (2). It is important to point out that the nonlinear eigenmodes $\varphi_{NL,n}(Z)$ and nonlinear eigenvalues $\beta_{NL,n}(Z)$ are $Z$-dependent (here $Z$ is the normalized propagation distance playing the role of time), because the nonlinear beam dynamics are generally not stationary. Indeed, we use a general theoretical protocol (developed recently in ref.[43]) for interpreting the dynamics in nonlinear topological systems, where both inherited and emergent topological phenomena may arise. The calculated nonlinear eigenvalue spectrum (at $Z = 50$) for the nontrivial SSH lattice is plotted in Fig. 2a, where two sets of edge modes set apart the whole band as in the linear spectrum (Fig. 1d1). However, under the action of nonlinearity, the spectrum exhibits a dynamical evolution during propagation, while the corner modes are forced to couple with the lower (upper) edge states by a self-focusing (self-defocusing) nonlinearity (see Supplementary movies). In other words, they are no longer stationary BICs, but rather undergo periodic energy exchange with the edge modes. The robustness of the corner localized BICs is evident, as they do not couple (for the parameters used here) with the bulk modes even when they are driven in and out of the central bulk band via nonlinearity, reflecting the topological nature of BICs. For the snapshot selected at $Z = 50$ shown in Fig. 2a, the whole spectrum is down-shifted from its linear position by the focusing nonlinearity, while the corner modes are approaching and coupling with the lower edge modes. This shifting direction is reversed when a self-defocusing nonlinearity is employed (see discussion in Supplementary material). When the strength of nonlinearity is low ($E_0' = 5, \gamma = 1.1$), a representative corner mode excited by the focusing nonlinearity undergoes beating with the edge modes (Figs. 2(b2, b4)), but only at sufficiently high nonlinearity ($E_0' = 5, \gamma = 3.5$) it is "liberated" from the continuum and turns into a self-trapped semi-infinite gap soliton (Fig. 2b3). Such corner solitons formed only in the strongly nonlinear regime with eigenvalues (i.e., propagation constants) residing beyond the lattice Bloch band, have been explored previously in the 2D square lattices[61,64] but not in the context of HOTIs.

In our experiments, we establish the nonlinear photonic HOTI platform by site-to-site writing of the 2D SSH lattices in a photorefractive crystal with a continuous-wave (CW) laser[65]. Experimental details can be found in Methods. For direct comparison, the lattices are written into three different structures (nontrivial SSH, square, and trivial SSH, as illustrated in the top panel of Fig. 3) by tuning the dimerization parameter, in accordance to Figs. 1(d1-d3), which in the experiment is achieved by controlling the intra-cell and inter-cell waveguide distances. We then excite the same corner site with a single Gaussian probe beam. Results obtained under *linear* excitation are shown in Fig. 3, where the probe beam itself has no nonlinear self-action but evolves into a characteristic corner state with non-zero intensity distribution at the two NNN sites along edges (Fig. 3a2), representing a typical topological BIC realized in the nontrivial SSH lattice. For all other cases of excitation, either at the edge and bulk of the nontrivial lattice, or at the same corner of the trivial lattices, the probe beam simply cannot be localized but instead displays strong radiation into the bulk/edge as shown in Figs. 3(a2-c2). To simulate such linear corner excitation, we set $I_P = 0$ in Eq. (1), and results obtained from numerical simulations in three different lattices are displayed in Figs. 3(a3-c3), which agree qualitatively well with the experimental observations.

We now discuss the experimental results pertinent to the *nonlinear* control of HOTI corner states as illustrated in Figs. 1(a-b) and analyzed in Fig. 2. Measurements of the output intensity profile of the probe beam under a corner excitation of the nontrivial lattice with both self-focusing and -defocusing nonlinearities after 20-mm propagation are shown in Fig. 4, where for reference Fig. 4a plots the linear output of the topological corner state. At low nonlinearity, a direct comparison with the linear output shows that the corner-localized state differs in this case from the topological corner mode, since now the energy goes to the second (nearest neighbor) and even the fourth sites along the edges [Figs. 4(b, e)]. This is because of the nonlinearity-induced coupling between the corner and edge modes, as illustrated in Fig. 1b. In our experiments, because the propagation distance set by the crystal length is typically smaller than the period of beating, we cannot observe a distinct beating oscillation between the corner and edge modes numerically shown in Fig. 2b4. At high self-focusing nonlinearity, the probe beam is localized again in the corner, forming a self-trapped semi-infinite gap corner soliton as shown in Fig. 4c[61,64], in agreement with what illustrated in Fig. 2b3. On the other hand, at a high defocusing nonlinearity, the corner excitation leads to strong spreading of the energy into the bulk as well as into the edges (Fig. 4f) due to nonlinear mode beating involving higher-band bulk states. These experimental results are corroborated by our numerical simulations based on the NLSE of Eq. (1) (see Supplementary Material for details), demonstrating clearly the concept of nonlinear control of the topological BICs in the HOTIs.

**Discussion**

The concept of topologically protected (higher-order) BICs, which is explored here under the action of nonlinearity, is somewhat an oxymoron. If two states are close in energy (or, in the optics language, the propagation constants are close), then it should be energetically inexpensive to couple these states. In a nonlinear HOTI system, topology is involved in the dynamics, and thus for the system studied here this common sense is questioned.

In our theoretical simulations and experiments, we have clearly interestingly found that topological higher-order BICs, which are corner states of our nonlinear 2D SSH lattice, are dominantly coupling to the edge states rather than to the bulk states. This happens despite the fact that the corner-localized BICs are embedded in the continuum of bulk states, gapped from the edge states, as clearly illustrated in Fig. 1d1. The weak self-focusing or -defocusing nonlinearity, for practically all excitations employed in this work, breaks the chiral symmetry and the crystalline symmetry of the lattice (the only exception is the excitation in Fig. 2, which preserves the $C_{4v}$ symmetry). By breaking these symmetries, nonlinearity in principle allows the corner states to couple with the bulk states of the continuum they are embedded in[53]. However, the overlap with the edge states induced by the nonlinearity is obviously much larger, which leads to the dominant coupling between the corner and the edge states. It should be noted that this behavior depends (also somewhat unintuitively) on the dimerization parameter $c$. If its magnitude increases (on the highly topologically nontrivial side), the gap between the topological corner BICs and the edge states increases (Fig. 1c), yet the dominant corner-edge coupling persists as opposed to the corner-bulk coupling.

A feature of interest for technological applications is the possibility to nonlinearly couple two corners; such coupling was proposed with exciton-polariton corner modes[48]. From the simulations presented in Fig. 2, it follows that such coupling should be possible in our lattice. Namely, if one excites a single corner, the initial state is then a superposition of four corner states that will beat; another view of the dynamics is that the nonlinearity will enable coupling to the edge states, and then to the adjacent corner. This type of dynamics is verified in our numerical simulation under proper initial conditions. It should be noted that in our experiments the coupling length between adjacent waveguides is not small compared to the length of the crystal along the propagation direction, such that it is not possible to observe the beating oscillations presented in Fig. 2. However, by using longer samples with different lengths, or using different platforms with stronger coupling, such beating should be observed.

The distinction between the discrete and the continuous models under large nonlinearities merits further discussion. The continuous NLSE of Eq. (1) offers a quantitatively better description of the experiment than the discrete model in Eq. (2) under the tight-binding approximation. However, it is well known when the two models start to deviate. For a linear lattice that is sufficiently deep, the discrete model is a good approximation of the dynamics; the parameters of the linear lattice employed here are in this regime. When the nonlinearity is weak, the lattice will not be strongly perturbed, and the discrete model is still a good approximation. However, for a large self-defocusing nonlinearity, the whole lattice structure at the excitation can be strongly deformed. For example, if a corner gets excited, a large self-defocusing nonlinearity significantly molds the corner area and enables coupling between the NNN sites and changes the nearest neighbor coupling as well, which is not captured by the discrete model of Eq. (2). In contrast, for a large self-focusing nonlinearity, the whole lattice structure is preserved despite the deep potential at the excitation site, and we found that the discrete model is still qualitatively accurate for the presented self-focusing dynamics.

Quite generally, for a weak nonlinearity and practically any excitation, the symmetries responsible for the nontrivial topology of the 2D SSH model are broken[53]. However, in a weakly nonlinear system, the topological properties can persist as they are *inherited* from the corresponding linear system[43]. This is the origin of the weak nonlinear coupling between the corner and the bulk modes discussed above. The fact that the topological properties are inherited is quantified and illustrated in Fig. 4d, showing the bulk polarizations $P_x$ and $P_y$ (related to the 2D Zak phase[15,17]) as a function of the parameter $c$ and the strength of the nonlinearity $\gamma_k$. The nonlinear system corresponding to Fig. 4d is the 2D SSH lattice with one out of four lattice sites in *all* unit cells excited, i.e., its on-site refractive index is changed in comparison to the other three sites in the unit cell. Even though this is a specific nonlinear excitation, it serves well to quantify how the topological feature is preserved after nonlinearity is introduced.

It is well known that, for the linear 2D SSH lattice, the polarizations are topologically quantized, $P_i = \frac{1}{2}$ for $c < 0$, and $P_i = 0$ otherwise. In the nonlinear case, the symmetry and topological protection are broken in the strict sense, however, we easily see from the illustrations that there is a sharp jump in the polarization as $c$ crosses zero, which is inherited from the topological phase transition occurring in the underlying linear system. We see that the jump signifying this phase transition is preserved in the nonlinear system as well. We expect that such inherited nonlinear topological properties exist also in HOTIs of the third-order or even higher-order formed in synthetic dimensions.

In conclusion, we have reported what we believe to be the first theoretical and experimental study of nonlinear BICs in HOTIs. Understanding the nonlinear topological phases is not only of fundamental interest, but it may also be crucial for the development of photonic devices based on topological corner modes, including HOTI lasers.

## Materials and Methods

**Experimental method for lattice writing and probing**

To demonstrate the scheme for the nonlinear control illustrated in Figs. 1(a, b), we employ a simple yet effective CW-laser writing technique[65] to establish the finite-sized photonic 2D SSH lattices with the desired structures shown in Figs. 3(a1-c1). The technique relies on writing the waveguides site-to-site in a 20mm-long nonlinear photorefractive (SBN:61) crystal. Different from the femtosecond laser writing method developed for glass materials[5], the SSH lattices written in the crystal can be readily reconfigurable in terms of lattice spacing and boundary structures. The experimental setup involves a low-power (up to 100mW) CW-laser beam ($\lambda$=532nm) to illuminate a spatial light modulator (SLM), which creates a quasi-non-diffracting writing beam with reconfigurable input positions. For the writing process, the modulated light beam (ordinarily-polarized) is used with self-focusing nonlinearity, but for the probing during the nonlinear control process, we use an extraordinarily-polarized Gaussian beam for lattice excitation with either self-focusing or –defocusing nonlinearity by switching the bias field direction[60,61]. Because of the noninstantaneous photorefractive nonlinearity, all waveguides remain intact during the writing, probing, and data acquisition period, except for the local index change due to the nonlinear self-action of the probe beam. Through a multi-step writing approach, the desired SSH lattices can be reconfigured from a nontrivial to a trivial structure by controlling the lattice spacing between the strong and weak bonds[63]. After the writing process is completed, the whole lattice structures can be examined by sending a set of Gaussian beams into the crystal to probe the waveguides one by one, then plot the superimposed outputs of the probe beam to get the lattice structure as shown in Figs. 3(a1-c1). Moreover, the probe beam can undergo either linear or nonlinear propagation through the lattice, depending on whether the bias field is turned on or not.

**Numerical methods for beam propagation simulation and Zak phase calculation**

The evolution of a light beam propagating in a photonic lattice is obtained by numerically solving Eq. (1) with the split-step Fourier technique, also referred to as the beam propagation method (BPM). The 2D SSH lattice structure has four lattice sites per unit cell (Fig. 1b), that is,

$$I_L(x, y) = \sum_{s=1}^{4} \sum_{i=0}^{N/4-1} \sum_{j=0}^{N/4-1} I_{L0} \exp\left(-\frac{(x-a_{sij})^2}{w_0^2/2} - \frac{(y-b_{sij})^2}{w_0^2/2}\right), \qquad (3)$$

where $(a_{1ij}, b_{1ij}) = (iT, jT)$, $(a_{2ij}, b_{2ij}) = (a + iT, jT)$, $(a_{3ij}, b_{3ij}) = (iT, a + jT)$, and $(a_{4ij}, b_{4ij}) = (a + iT, a + jT)$, with $T = a + b$ being the lattice period, and $a$ and $b$ being the spacing between lattice sites for the weak and strong bonds (corresponding to intracell and intercell coupling in Fig. 1b, respectively). The total number of unit cells is $N^2/4$, where $w_0$ is a scaling factor and $I_{L0}$ is the lattice magnitude. In the experiments, depending on the relative values between $a$ and $b$, the photonic lattice can be reconfigured into a simple square lattice ($a = T/2$), a nontrivial SSH lattice for $a > T/2$ and a trivial SSH lattice for $a < T/2$. Similarly, we numerically excite only one corner (the left one in Fig. 3) with a Gaussian beam and perform the BPM simulation for subsequent dynamics under linear and nonlinear conditions. Linear propagation results obtained with different lattice parameters are illustrated in Figs. 3(a3-c3). For the nonlinear regime ($I_P \neq 0$), simulations are performed for both self-focusing and -defocusing nonlinearities in the nontrivial SSH lattice, and the results obtained at low and high nonlinearity are in good agreement with experimental observations (see Supplementary Material).

To characterize the topological properties of the 2D SSH lattices, we calculate the topological invariant based on the 2D polarization, which is defined for an infinite periodic system as[17,18]

$$P_i = -\frac{1}{(2\pi)^2} \iint dk_x dk_y Tr[A_i(k_x, k_y)], \qquad (4)$$

where $i = x, y$, $(A_i)_{mn}(\mathbf{k}) = i\langle u_m(\mathbf{k})|\partial_{k_i}|u_n(\mathbf{k})\rangle$ is the Berry connection, and $u_m(\mathbf{k})$ is the eigenmode in the $m^{th}$ band. The 2D polarization is directly related to the 2D Zak phase: $Z_i = 2\pi P_i$. One can readily calculate the polarization in the linear regime, which yields $P_x = P_y = \frac{1}{2}$ for $c < 0$, and $P_x = P_y = 0$ for $c > 0$.

In order to test whether the signature of the topological phase transition at $c = 0$ is still present in the nonlinear regime, we calculate the nonlinear polarization by employing Eq. (4) for the following modified Hamiltonian for the 2D SSH lattices:

$$\hat{H} = \begin{pmatrix} \gamma_k & t + t'\exp(-ik_x) & t + t'\exp(-ik_y) & 0 \\ t + t'\exp(ik_x) & 0 & 0 & t + t'\exp(-ik_y) \\ t + t'\exp(ik_y) & 0 & 0 & t + t'\exp(-ik_x) \\ 0 & t + t'\exp(ik_y) & t + t'\exp(ik_x) & 0 \end{pmatrix}, \quad (5)$$

where $\gamma_k$ accounts for the nonlinearity strength, and its sign manifests the difference between self-focusing and -defocusing nonlinearities. This Hamiltonian corresponds to exciting one out of four lattice sites in all unit cells and changing its on-site potential via the employed nonlinearity. Calculated results for the nonlinear polarization are plotted in Fig. 4d, as a function of the dimerization parameter $c = t - t'$ defined earlier.

**Acknowledgments**

This research is supported by the National Key R&D Program of China under Grant No. 2017YFA0303800, the National Natural Science Foundation (11922408, 91750204, 11674180), PCSIRT, and the 111 Project (No. B07013) in China. D. B. acknowledges support from 66 Postdoctoral Science Grant of China. D.J. and H.B. acknowledge support in part by the Croatian Science Foundation Grant No. IP-2016-06-5885 SynthMagIA, and the QuantiXLie Center of Excellence, a project co-financed by the Croatian Government and European Union through the European Regional Development Fund - the Competitiveness and Cohesion Operational Programme (Grant KK.01.1.1.01.0004). R.M. gratefully acknowledges support from the NSERC Discovery and Canada Research Chair Programs. R.M. is affiliated to 6 as an adjoint professor.

**Conflict of interests**

The authors declare no conflicts of interest. The authors declare no competing financial interests.

**Contributions**

All authors contributed to this work.

Supplementary information accompanies the manuscript is available on the Light: Science & Applications website (http://www.nature.com/lsa).

Correspondence and requests for materials should be addressed to Z.C. or H.B.

Note: We became aware of a relevant work when this paper was finalized for submission: https://arxiv.org/abs/2104.13112

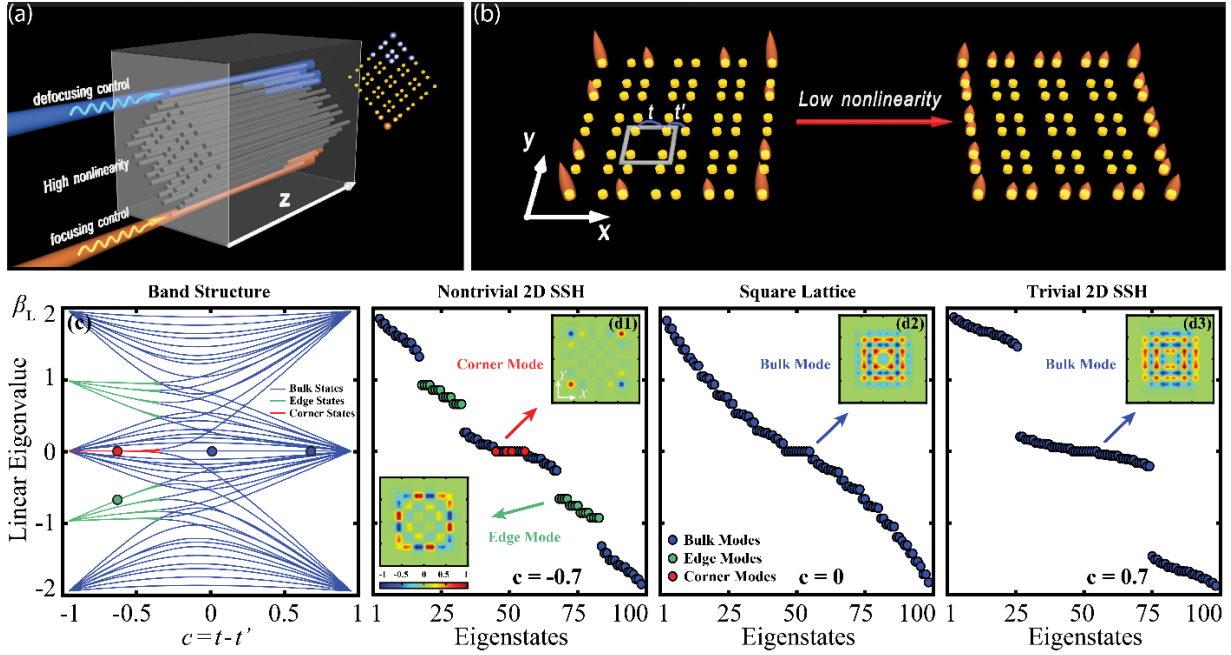

**Figure 1. Illustration of nonlinear control of a higher-order topological insulator.** (a) Schematic of corner excitations in a 2D SSH photonic lattice under high nonlinearity, where a focusing nonlinearity leads to corner soliton formation while a defocusing nonlinearity leads to radiation into the bulk/edge. (b) Coupling and beating between corner and edge states under weak nonlinearity. (c) Calculated *linear* eigenvalues of the SSH lattice $\beta_L$ as a function of the dimerization parameter $c$, where the corner and edge states are highlighted with red and green colors in the highly topologically nontrivial regime. (d1-d3) Calculated band structures for the nontrivial, square and trivial lattices, showing the topological phase transition as the dimerization parameter is tuned, where the insets plot the selected mode profiles corresponding to the marked color points. A topological BIC with characteristic corner-localized mode profile is shown in the upper-right inset, with zero amplitude in the nearest neighboring sites but nonzero amplitude and opposite phase in the NNN sites.

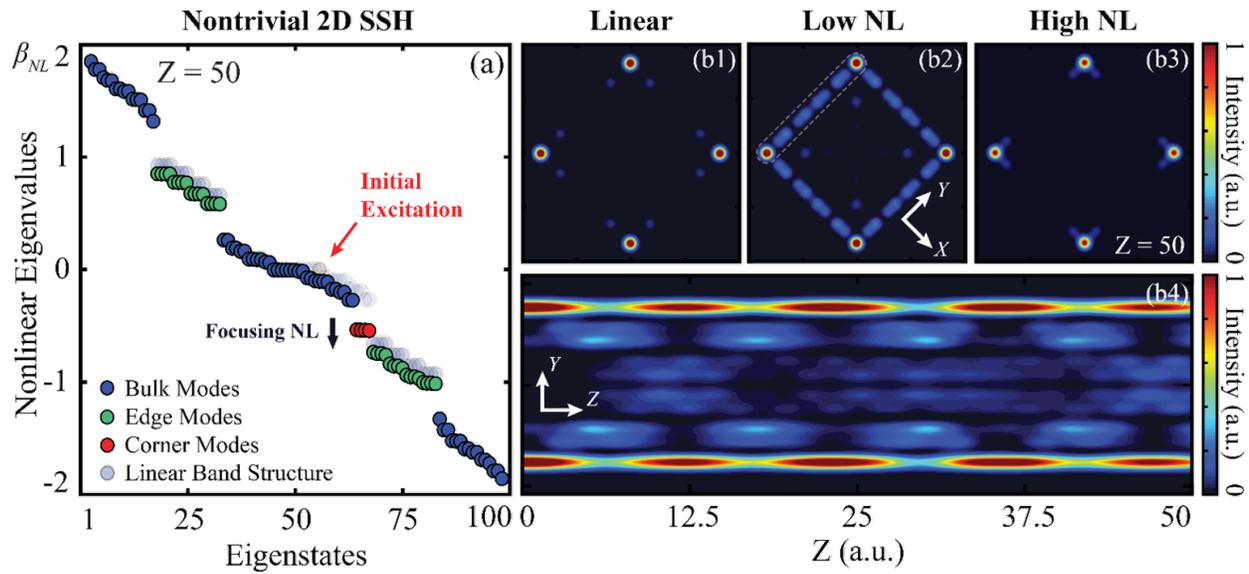

**Figure 2. Calculated nonlinear band structure and corner mode tuning under self-focusing nonlinearity.** (a) Calculated *nonlinear* eigenvalues of the SSH lattice $\beta_{NL}$ for the nontrivial lattice using the discrete model, where the transparent dots are linear modes superimposed for direct comparison. The black arrow illustrates that four corner states (red dots) undergo coupling and beating with lower edge states under low self-focusing nonlinearity (see also Supplementary video), and the red arrow marks the initially excited corner mode which sustains the topological feature under linear condition as shown in (b1) without any light distribution in the nearest neighboring sites. Under a low focusing nonlinearity, the corner mode couples with the edge modes (b2), and a beating oscillation occurs. This can be seen clearly from the side-view propagation of (b4), taking from the upper-left edge marked by a dashed line in (b2). Under a high focusing nonlinearity, a localized semi-infinite gap discrete soliton forms at the corners, with evident light distribution in the nearest neighboring sites (b3).

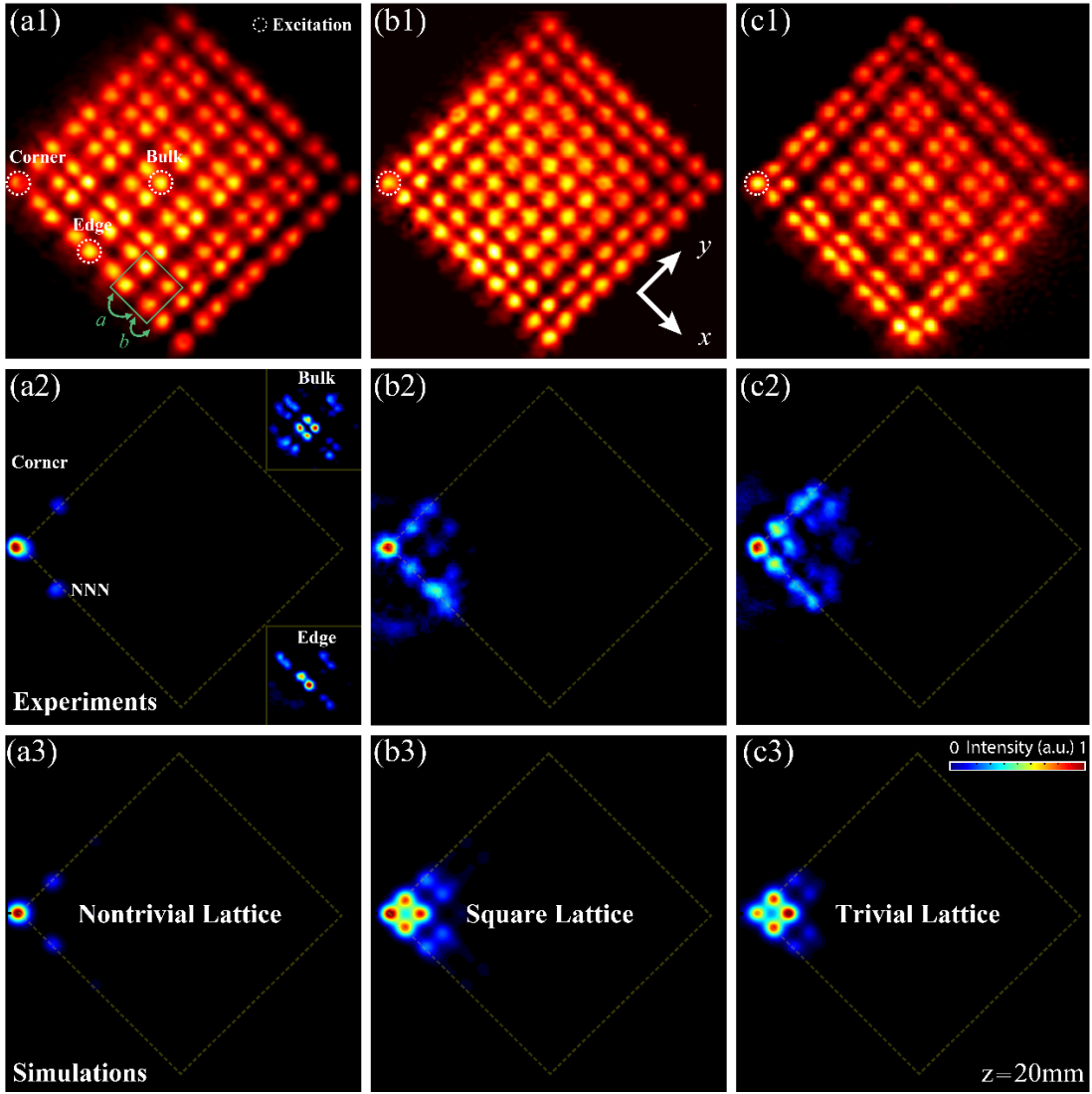

**Figure 3. Experimental realization and probing of linear 2D photonic SSH lattices**. (a1-c1) Laser-written 2D SSH lattices tuned to nontrivial, square and trivial regimes at different dimerizations, where the dashed circles indicate the lattice sites for corner, edge, and bulk excitations, and *a* and *b* mark the waveguide distances for the weak and strong bonds. (a2-c2) Experimental results of linear output corresponding to single-site excitations in (a1-c1), where the corner excitation in (a1) leads to a localized BIC with evident topological features: no light distribution in the nearest neighboring sites but a non-zero intensity in the NNN sites along two edges. Discrete diffraction is observed for all other excitations. (a3-c3) Numerical results corresponding to corner excitations in (a2-c2) obtained using the continuum model, where the propagation distance is 20 mm corresponding to the length of the crystal used in the experiments. Experimental parameters: $a = 31 \mu m, b = 23 \mu m$; the bias field during writing process is $E_0 = 130 kV/m$.

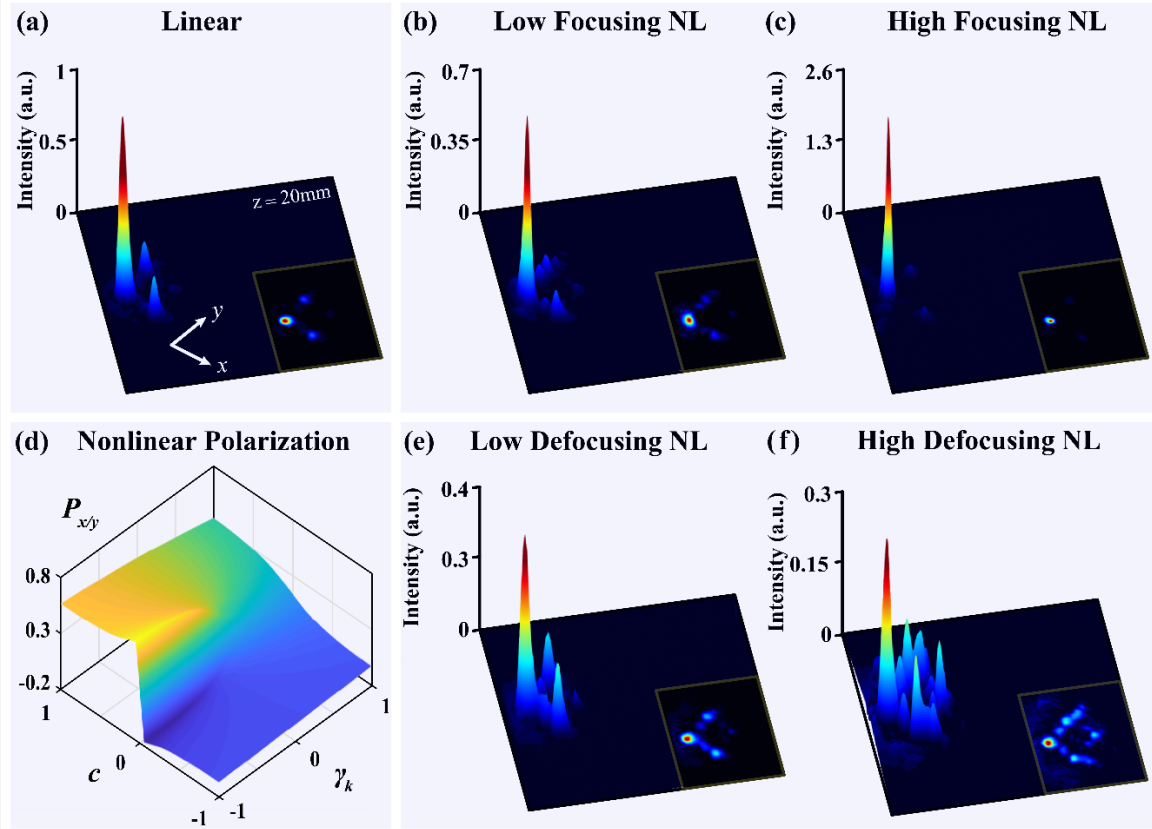

**Figure 4. Experimental demonstration of nonlinear control of a higher-order topological insulator.** (a) 3D-view of a typical linear corner state experimentally observed in a nontrivial lattice. (b, c) Nonlinear self-focusing leads to (b) coupling into the edges (non-zero intensity along the edge sites compared with linear case) when the nonlinearity is low, and (c) a highly localized corner soliton when the nonlinearity is high. (d) Plot of the calculated nonlinear polarization as a function of the nonlinear control parameter $\gamma_k$ as well as the dimerization parameter $c$. Characteristic jump in the bulk polarization manifesting the topological phase transition extends beyond the linear condition ($\gamma_k = 0$) because of the inherited topology under the nonlinear condition. (e, f) Experimental results of nonlinear control with a low and high self-defocusing nonlinearity. Under a high defocusing nonlinearity, the energy spreads dramatically to both the edge and the bulk (f). For the focusing (defocusing) case, the bias filed is $E_0 = 160kV/m$ ($E_0 = -80kV/m$), and the average power of the probe beam is about $15nW$ ($70nW$) for the low (high) nonlinearity. See Supplementary material for corresponding numerical simulation results.